\documentclass[prl,showpacs,twocolumn,preprintnumbers]{revtex4}
\usepackage{mathptmx}
\usepackage[english]{babel}
\usepackage{amssymb}
\usepackage{amsfonts}
\usepackage{amsmath}
\usepackage{eucal}
\usepackage{graphicx}
\usepackage{epsfig}
\newcommand{\cA}{{\cal A}}  \newcommand{\cB}{{\cal B}}

\newcommand{\cI}{{\cal I}}

  \newcommand{\cP}{{\cal P}}
  \newcommand{\cR}{{\cal R}}
  
  \newcommand{\cV}{{\cal V}}

\newcommand{\be}{\begin{equation}} \newcommand{\ee}{\end{equation}}
\newcommand{\bea}{\begin{eqnarray}} \newcommand{\eea}{\end{eqnarray}}
\newcommand{\beann}{\begin{eqnarray*}}  \newcommand{\eeann}{\end{eqnarray*}}
\newcommand{\bfig}{\begin{figure}} \newcommand{\efig}{\end{figure}}
\newcommand{\ba}{\begin{array}} \newcommand{\ea}{\end{array}}
\newcommand{\bcen}{\begin{center}} \newcommand{\ecen}{\end{center}}
\newcommand{\btab}{\begin{tabular}} \newcommand{\etab}{\end{tabular}}

\def\tr{\operatorname{tr\:}}     
   
     \def\diag{\operatorname{diag}}

\newtheorem{Proposition}{Proposition}[section]

\newtheorem{Theorem}{Theorem}[section]
\newtheorem{Lemma}{Lemma}[section]
\newtheorem{Corrolary}{Corrolary}[section]

\newcommand{\bp}{\begin{Proposition}}	\newcommand{\ep}{\end{Proposition}}
\newcommand{\bt}{\begin{Theorem}}	\newcommand{\et}{\end{Theorem}}
\newcommand{\bl}{\begin{Lemma}}		\newcommand{\el}{\end{Lemma}}
\newcommand{\bc}{\begin{Corrolary}}	\newcommand{\ec}{\end{Corrolary}}
\begin{document}

\title{Gravitational anomaly and transport}

\author{Karl Landsteiner}
\author{Eugenio Meg\'{\i}as}
\author{Francisco Pena-Benitez}
\affiliation{Instituto de F\'{\i}sica Te\'orica UAM/CSIC, C/ Nicol\'as Cabrera
13-15,\\
Universidad Aut\'onoma de Madrid, Cantoblanco, 28049 Madrid, Spain}

\begin{abstract}
Quantum anomalies give rise to new transport phenomena. In particular
a magnetic field can induce an anomalous current via the chiral
magnetic effect and a vortex in the relativistic fluid can
also induce a current via the chiral vortical effect.  The related transport
coefficients can be calculated via Kubo formulas. We evaluate the Kubo
formula for the anomalous vortical conductivity at weak coupling and
show that it receives contributions proportional to the gravitational
anomaly coefficient. The gravitational anomaly gives rise to an anomalous
vortical
effect even for an uncharged fluid.
\end{abstract}

\pacs{11.15-q, 11.40-q, 11.10.Wx, 67.10.Jn}
\preprint{IFT-UAM/CSIC-11-15}
\preprint{ESI 2328}
\maketitle
%
{\bf Introduction.}
The effects quantum anomalies have on the hydrodynamics of otherwise
conserved currents have recently received much attention.  Two such
effects are known: an anomalous magnetic effect
\cite{Fukushima:2008xe} and an anomalous vortical effect
\cite{Kharzeev:2007tn,vorticity}. In the first an (external) magnetic field
induces a
current via the axial anomaly whereas the second is the generation of
a current due to the presence of a vortex in the charged relativistic
fluid.  These effects have been argued to lead to observable event by
event parity violation and a charge separation effect in non-central
heavy ion collision at RHIC and LHC \cite{experiments}.  In the hydrodynamic
constitutive
relations these effects lead to the existence of a new class of
transport coefficients.

A first principle calculation of transport coefficients is possible
via so called Kubo formulas. 
The Kubo formula for the anomalous magnetic conductivity has
been derived and applied in \cite{Kharzeev:2009pj} whereas the one for the
anomalous vortical conductivity has been established recently in
\cite{Amado:2011zx}.
They are
\begin{eqnarray}
\sigma^\cB_{AB} &=& \lim_{k_n\rightarrow 0} \sum_{ij}\epsilon_{ijn}
\frac{-i}{2k_n} \langle J_A^i J_B^j \rangle|_{\omega=0} \, ,\label{eq:sigmaB}\\
\sigma^\cV_{A}  &=& \lim_{k_n\rightarrow 0} \sum_{ij}\epsilon_{ijn}
\frac{-i}{2k_n} \langle J_A^i T^{0j} \rangle|_{\omega=0} \, ,
 \label{eq:sigmaV} 
 \end{eqnarray}
where $J_A^\mu$ are the (anomalous) currents and $T^{\mu\nu}$ is the energy
momentum tensor. 
These Kubo formulas allow the calculation of the transport coefficients in the
constitutive relations
\begin{eqnarray}
J_A^i &=& \sigma^\cB_{AB}\epsilon^{ijk}\partial_j \cA^B_k +
\sigma^\cV_{A}\epsilon^{ijk}\partial_j v_k\,, 
\label{eq:JiA}
\end{eqnarray}
where 
$\cA^B_k$ are the spatial components of a collection of gauge fields and $v_k$
is the local fluid velocity. The fluid velocity is taken here in a
laboratory frame, i.e. we take $v_k$ independent of the external magnetic
fields. There will also be a non-zero energy flow
which could be compensated for by going to the Landau frame. Since we are
interested only in the total current generated by
the external magnetic fields we will not do so. See also \cite{Amado:2011zx} for
a discussion. 
We note that we can substitute the fluid velocities by the gravitomagnetic
potential. To do so we go to the rest frame
of the fluid defined by $u^\mu=(1,0,0,0)$ but switch on a gravitomagnetic field
in the metric according to
\begin{equation}
ds^2 = dt^2 + 2 \vec{\cA_g}d\vec{x} dt - d\vec{x}^2\,.
\end{equation}
Using this metric we can compute the local fluid velocity $u_\mu =
(1,\vec{\cA}_g)$ such that $\vec{v}=\vec{\cA_g}$. 
All these expressions are valid only up to first order in the external fields
$\cA^A_k, v_k$. Plugging this into the constitutive relation (\ref{eq:JiA}) and
noting that $\cA^i_g$ sources $T^{0i}$ leads to the
Kubo formula for the vortical conductivity.

We will now evaluate the Kubo formulas (\ref{eq:sigmaB}),
(\ref{eq:sigmaV}) 
for a
theory of $N$ free right-handed fermions $\Psi^f$ transforming under a
global symmetry group $G$ generated by matrices $(T_A)^f\,_g$.  We
denote the generators in the Cartan subalgebra by $H_A$. Chemical
potentials $\mu_A$ can be switched on only in the Cartan
subalgebra. Furthermore the presence of the chemical potentials breaks
the group $G$ to a subgroup $\hat G$. Only the currents that lie in
the unbroken subgroup are conserved (up to anomalies) and participate
in the hydrodynamics. The chemical potential for the fermion $\Psi^f$
is given by $\mu^f= \sum_A q_A^f  \mu_A$, where we write the Cartan generator
$H_A = q_A^f\delta^f\,_g$ in terms of its eigenvalues, the charges
$q_A^f$. The unbroken symmetry group $\hat G$ is generated by the
matrices $T_A^f\,_g$ fulfilling
\begin{equation}\label{eq:unbroken}
 T_A^f\,_g \mu^g = \mu^f T_A^f\,_g\,.
\end{equation}
There is no summation over indices in the last expression. From now on we will
assume that all currents $\vec{J}_A$ lie in directions indicated in 
(\ref{eq:unbroken}). We define the chemical potential through boundary
conditions on the fermion fields around the thermal circle
\cite{Landsman:1986uw}, $ \Psi^f(\tau) = - e^{\beta \mu^f} \Psi^f(\tau-\beta)$
with $\beta=1/T$. 
Therefore the eigenvalues of $\partial_\tau$ are
$i\tilde\omega_n+\mu^f$ for the fermion species $f$ with
$\tilde\omega_n=\pi T(2n+1)$ the fermionic Matsubara frequencies.  A
convenient way of expressing the currents is in terms of Dirac fermions
and writing
\begin{eqnarray}
J^i_A &=& \sum_{f,g=1}^N T_A^g\,_f \bar\Psi_g \gamma^i \cP_+ \Psi^f \,,
\label{eq:JA}\\
T^{0i} &=&  \frac i 2 \sum_{f=1}^N\bar\Psi_f  ( \gamma^0  \partial^i + \gamma^i
\partial^0  ) \cP_+\Psi^f\,, \label{eq:JE}
\end{eqnarray}
where we used the chiral projector $\cP_\pm = \frac 1 2 (1\pm\gamma_5)$.
The fermion propagator is
\begin{eqnarray}
S(q)^f\,_g &=&  \frac{\delta^f\,_g}{2} \sum_{t=\pm}
\Delta_t(i\tilde\omega^f,\vec{q}) \cP_+ \gamma_\mu \hat q^\mu_t \,,\\
\Delta_t( i\tilde\omega^f, q) &=& \frac{1}{i\tilde\omega^f - t E_q}\,,
\end{eqnarray}
with  $i\tilde\omega^f = i\tilde\omega_n + \mu^f$, $\hat q_t^\mu = (1, t \hat
q)$, $\hat{q} = \frac{\vec{q}}{E_q}$ and $E_q=|\vec q |$. 
We can easily include left-handed fermions as well. They contribute in all our
calculations in the same way as the right handed ones up to a relative minus
sign.\\

{\bf Evaluation of Kubo formulas.}
We will address in detail the computation of the vortical conductivities
Eq.~(\ref{eq:sigmaV}) and sketch only the calculation
of the magnetic conductivities since the latter one is a trivial extension of
the calculation of the chiral magnetic conductivity in \cite{Kharzeev:2009pj}.\\

\begin{center}
\begin{figure}[tbp]
\epsfig{figure=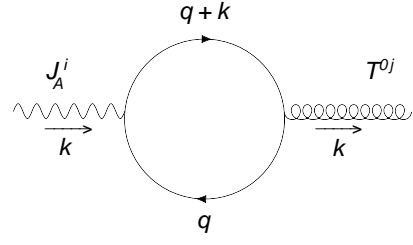,scale=1}
\caption{1 loop diagram contributing to the vortical conductivity
Eq.~(\ref{eq:sigmaV}).}
\label{fig:1loop}
\end{figure}
\end{center}

{\it Vortical conductivity:}
 The vortical conductivity is defined from the retarded correlation
 function of the current $J^i_A(x)$~(\ref{eq:JA}), and the
 energy momentum tensor or energy current  $T^{0j}(x^\prime)$~(\ref{eq:JE}),
i.e.
\begin{equation}
G_A^\cV(x-x^\prime) = \frac{1}{2} \epsilon_{ijn}\,i \, \theta(t-t^\prime) 
\,\langle [J^i_{A}(x),T^{0j}(x^\prime)] \rangle \,.
\end{equation}
Going to Fourier space, one can evaluate this quantity as
\begin{eqnarray}
G_A^\cV(k) &=& \frac{1}{4}\sum_{f=1}^N T_A^f\,_f
\frac{1}{\beta}\sum_{\tilde\omega^f} \int\frac{d^3q}{(2\pi)^3} \epsilon_{ijn}
\tr \Bigg[ S^f\,_f(q) \gamma^i \nonumber \\
&&\qquad\times\; S^f\,_f(q+k) \left(  \gamma^0 q^j + \gamma^j i\tilde\omega^f
\right)  \Bigg] \,, \label{eq:Gav}
\end{eqnarray}
which corresponds to the one loop diagram of Fig.~\ref{fig:1loop}. 

The vertex of the two quarks with the graviton is $\sim \delta^f\,_g$, and
therefore we find only contributions from the diagonal part of the group~$\hat
G$. Our metric is $g_{\mu\nu} = \diag(1,-1,-1,-1)$. We can split $G_A^\cV$ into
two contributions
\begin{equation}
G_A^\cV(k) = G_{A,(0j)}^\cV(k) + G_{A,(j0)}^\cV(k) \,, \label{eq:Gav2}
\end{equation}
which correspond to the terms $\gamma^0 q^j$  and $\gamma^j i\tilde\omega^f$ in
Eq.~(\ref{eq:Gav}) respectively. We will focus first on the computation of
$G_{A,(0j)}^\cV$. The integrand of Eq.~(\ref{eq:Gav}) for $G_{A,(0j)}^\cV$ can
be written as
\begin{eqnarray}
 \cI_{A,(0j)}^\cV &=& \frac{1}{4}q^j\sum_{t,u=\pm} \epsilon_{ijn} \tr[\gamma_\mu
\gamma^i \gamma_\nu \gamma^0 \cP_+ ] \times \label{eq:integrand0j} \\
&&\times \Delta_t(i\tilde\omega^f,\vec{q}) \Delta_u(i\tilde\omega^f +
i\omega_n,\vec{q}+\vec{k}) {\hat q}^\mu_t {\widehat{(q+k)}}^\nu_u \,. 
\nonumber 
\end{eqnarray} 

From a computation of the Dirac trace in Eq.~(\ref{eq:integrand0j}) one has two
contributions
\begin{eqnarray}
&&\epsilon_{ijn} \tr[\gamma_\mu \gamma^i \gamma_\nu \gamma^0] a^\mu b^\nu = 4
\epsilon_{ijn} ( a^i b^0 + a^0 b^i ) \,, \label{eq:trace1} \\
&&\epsilon_{ijn} \tr[\gamma_\mu \gamma^i \gamma_\nu \gamma^0 \gamma_5] a^\mu
b^\nu = 4i(a_j b_n - a_n b_j) \,.  \label{eq:trace2}
\end{eqnarray}
Using Eqs.~(\ref{eq:integrand0j}), (\ref{eq:trace1}) and (\ref{eq:trace2}) one
can express $G_{A,(0j)}^\cV(k)$ as
\begin{eqnarray}
G_{A,(0j)}^\cV(k) &=& \frac{1}{8} \sum_{f=1}^N  T_A^f\,_f
\frac{1}{\beta}\sum_{\tilde\omega^f} \int \frac{d^3q}{(2\pi)^3}   q^j
\sum_{t,u=\pm} \Bigg[  \epsilon_{ijn} \Bigg( t\frac{q^i}{E_q} \nonumber \\
&+& u \frac{k^i + q^i}{E_{q+k}}\Bigg) + i\frac{tu}{E_q E_{q+k}} \left( q_j k_n -
q_n k_j  \right) \Bigg] \nonumber \\
&\times&  \Delta_t(i\tilde\omega^f,\vec{k})
\Delta_u(i\tilde\omega^f+i\omega_n,\vec{q}+\vec{k})  \,. \label{eq:Gav0j2}
\end{eqnarray}
At this point one can make a few simplifications. Note that due to the
antisymmetric tensor $\epsilon_{ijn}$, the two terms proportional to $q^i$
inside the bracket in Eq.~(\ref{eq:Gav0j2}) vanish. Regarding the term 
$\epsilon_{ijn} q^j k^i$, it leads to a contribution $\sim \epsilon_{ijn} k^j
k^i$ after integration in $d^3q$, which is zero. Then the only term which
remains is the one not involving $\epsilon_{ijn}$. We can now perform the sum
over fermionic Matsubara frequencies. One has
\begin{eqnarray}
&&\frac{1}{\beta}\sum_{\tilde\omega^f} \Delta_t(i\tilde\omega^f,\vec{q})
\Delta_u(i\tilde\omega^f+i\omega_n,\vec{q}+\vec{p}) = \nonumber  \\
&&\quad= \frac{t n(E_q - t\mu^f) - u n(E_{q+k}-u \mu^f) +
\frac{1}{2}(u-t)}{i\omega_n + t E_q - u E_{q+k}} \,, \label{eq:Matsubara1}
\end{eqnarray}
where $n(x) = 1/(e^{\beta x} + 1)$ is the Fermi-Dirac distribution function. In
Eq.~(\ref{eq:Matsubara1}) we have considered that $\omega_n = 2\pi T n$ is a
bosonic Matsubara frequency. This result is also obtained in
Ref.~\cite{Kharzeev:2009pj}. After doing the analytic continuation, which
amounts to replacing $i\omega_n$ by $k_0 + i\epsilon$ in
Eq.~(\ref{eq:Matsubara1}), one gets
\begin{eqnarray}
&&G_{A,(0j)}^\cV(k) = -\frac{i}{8} \sum_{f=1}^N T_A^f\,_f \int
\frac{d^3q}{(2\pi)^3}  \frac{ \vec{q}^2 k_n - (\vec{q}\cdot\vec{k}) q_n }{E_q
E_{q+k}}  \label{eq:Gav0j3} \\ 
&&\quad\times \sum_{t,u=\pm}  \frac{u n(E_q - t\mu^f) - t n(E_{q+k}-u \mu^f) +
\frac{1}{2}(t-u)}{k_0 + i\epsilon + t E_q - u E_{q+k}}\,. \nonumber
\end{eqnarray} 
The term proportional to~$\sim \frac{1}{2}(t-u)$ corresponds to the vacuum
contribution, and it is ultraviolet divergent. By removing this term the finite
temperature and chemical potential behavior is not affected, and the result
becomes ultraviolet finite because the Fermi-Dirac distribution function
exponentially suppresses high momenta. By making both the change of variable
$\vec{q} \to -\vec{q} -\vec{k}$ and the interchange $u \to -t$ and $t \to -u$ in
the part of the integrand involving the term $-tn(E_{q+k}-u\mu^f)$, one can
express the vacuum subtracted contribution of Eq.~(\ref{eq:Gav0j3}) as
\begin{eqnarray}
\widehat{G}_{A,(0j)}^\cV(k) &=& \frac{i}{8} k_n \sum_{f=1}^N T_A^f\,_f \int
\frac{d^3 q}{(2\pi)^3} \frac{1}{E_q E_{q+k}} \left( \vec{q}^2 -
\frac{(\vec{q}\cdot\vec{k})^2}{\vec{k}^2}\right)  \nonumber \\
&&\times\sum_{t,u=\pm} u \frac{n(E_q - \mu^f) + n(E_q+ \mu^f)}{k_0 + i\epsilon +
t E_q + u E_{q+k}}\,. \label{eq:Gav0j4}
\end{eqnarray}
The result has to be proportional to $k_n$, so to reach this expression we have
replaced $q_n$ by $(\vec{q}\cdot \vec{k})k_n/\vec{k}^2$ in
Eq.~(\ref{eq:Gav0j3}). At this point one can perform the sum over $u$ 
and the integration over angles. 
Then one gets the final result
\begin{eqnarray}
&&\widehat{G}_{A,(0j)}^\cV(k) = \frac{i}{16\pi^2}\frac{k_n}{k^2}(k^2 - k_0^2)
\sum_{f=1}^N T_A^f\,_f \int_0^\infty dq \, q \, f^\cV(q)\Bigg[ 1 \nonumber \\
&&+ \frac{1}{8qk} \sum_{t=\pm} \left[k_0^2-k^2+4q(q+t k_0)\right] \log \left(
\frac{\Omega_t^2 - (q+k)^2}{\Omega_t^2 - (q-k)^2}\right) \Bigg] \,, \nonumber
\\ 
\label{eq:Gav0j5} 
\end{eqnarray} 
where~$\Omega_t = k_0 + i\epsilon + t E_q\,,$ and
$
f^\cV(q) = n(E_q-\mu^f) + n(E_q + \mu^f) \,.   \label{eq:fV}
$
The steps to compute $G_{A,(j0)}^\cV$ in Eq.~(\ref{eq:Gav2}) are similar. In
this case the Dirac trace leads to a different tensor structure, in  which the
only contribution comes from the trace involving~$\gamma_5$, i.e.
\begin{equation}
\epsilon_{ijn} \tr[\gamma_\mu \gamma^i \gamma_\nu \gamma^j \gamma_5] a^\mu b^\nu
=8i ( a_n b_0 - a_0 b_n ) \,. \label{eq:trace3}
\end{equation}
The sum over fermionic Matsubara frequencies involves an
extra~$i\tilde\omega^f$. Following the same procedure as explained above, the
vacuum subtracted contribution writes
\begin{eqnarray}
&&\widehat{G}_{A,(j0)}^\cV(k) = -\frac{i}{32\pi^2} \frac{k_n}{k^3}\sum_{f=1}^N
T_A^f\,_f \int_0^\infty dq \sum_{t=\pm} f^\cV_t(q,k_0) \label{eq:Gavj05} \\
&&\times \Bigg[  4 t q k k_0  -\left(k^2-k_0^2\right) (2q + t k_0) \log\left(
\frac{\Omega_t^2  - (q+k)^2}{\Omega_t^2 - (q-k)^2 } \right) \Bigg] \,, \nonumber
\end{eqnarray}
where
$
f^\cV_t(q,k_0) = q f^\cV(q) + t k_0 n(E_q + t\mu^f)   \,. \label{eq:fVt}
$
The result for $\hat{G}_A^\cV(k)$ writes as a sum of Eqs.~(\ref{eq:Gav0j5}) and
(\ref{eq:Gavj05}), according to Eq.~(\ref{eq:Gav2}). From these expressions one
can compute the zero frequency, zero momentum, limit. Since
\begin{equation}
\lim_{k\to 0} \lim_{k_0 \to 0} \sum_{t=\pm}\log \left( \frac{\Omega_t^2 -
(q+k)^2}{\Omega_t^2 - (q-k)^2}\right) = \frac{2k}{q} \,, \label{eq:limlog}
\end{equation}
the relevant integrals are
\begin{equation}
\int_0^\infty dq \, q \, f^\cV(q) = \int_0^\infty dq \, f^\cV_t(q,k_0 = 0) =
\frac{(\mu^f)^2}{2} + \frac{\pi^2}{6} T^2 \,.
\end{equation}
Finally it follows from Eqs.~(\ref{eq:Gav0j5}) and (\ref{eq:Gavj05}) that the
zero frequency, zero momentum, vortical conductivity writes
\begin{eqnarray}
\sigma_A^\cV &=& \frac{1}{8\pi^2} \sum_{f=1}^N T_A^f\,_f \Bigg[ ( \mu^f)^2 +
\frac{\pi^2}{3} T^2 \Bigg] \label{eq:sigmaV2} \\
&=& \frac{1}{16\pi^2}  \Bigg[ \sum_{B,C} \tr\left(T_A \{ H_B , H_C \}\right)
\mu_B \, \mu_C  + \frac{2\pi^2}{3} T^2 \tr \left( T_A \right) \Bigg]  \,.
\nonumber
\end{eqnarray}
Both $\widehat{G}_{A,(0j)}^\cV$ and $\widehat{G}_{A,(j0)}^\cV$ lead to the same
contribution in~$\sigma_A^\cV$. Eq.~(\ref{eq:sigmaV2}) constitutes our main
result. The term involving the chemical potentials is induced by the chiral
anomaly. More interesting is the term $\sim T^2$ which is proportional to the
gravitational anomaly  \cite{GravAnom} as we will show in the discussion section. The Matsubara frequencies $\tilde\omega_n=\pi T (2n+1)$ generate a dependence on $\pi T$ in the final result as compared to the chemical potentials, and then no factors of $\pi$ show up for the term $\sim T^2$ in Eq.~(\ref{eq:sigmaV2}). Left handed fermions contribute in the same way but with a relative minus sign.
Left and right handed fermions do not mix.

If instead of having taken the zero momentum limit at zero frequency, one took
the zero frequency limit at zero momentum, the result would be $1/3$ of the
result quoted in Eq.~(\ref{eq:sigmaV2}). The same factor appears in the magnetic
conductivity when one interchanges the two limits~\cite{Kharzeev:2009pj}.\\

{\it Magnetic conductivity:}
The magnetic conductivity in the case of a vector and an axial $U(1)$ symmetry
was computed at weak coupling in~\cite{Kharzeev:2009pj}. Following the same
method, we have computed it for the unbroken (non-abelian) symmetry group~$\hat
G$.
The relevant Green function is 
\begin{eqnarray}
G^\cB_{AB} &=& \frac 1 2 \sum_{f,g} T_A^g\,_f T_B^f\,_g \frac 1 \beta 
\sum_{\tilde\omega^f} \int\frac{d^3q}{(2\pi)^3} \epsilon_{ijn} \nonumber \\
&&\qquad\qquad\times \tr \Bigg[ S^f\,_f(q) \gamma^i S^f\,_f(q+k) \gamma^j   
\Bigg] \,. \label{eq:Gabb}
\end{eqnarray}
The evaluation of this expression is exactly as in \cite{Kharzeev:2009pj} so we
skip the details. 
 The result is 
\begin{equation}
\sigma_{AB}^\cB =  \frac{1}{4\pi^2} \sum_{f,g=1}^N T_A^f\,_g T_B^g\,_f \, \mu^f
= \frac{1}{8\pi^2} \sum_C \tr\left( T_A \{ T_B , H_C \} \right) \, \mu_C \,.
\label{eq:sigmaB1}
\end{equation}
In the second equality of Eq.~(\ref{eq:sigmaB1}) we have made use of
Eq.~(\ref{eq:unbroken}). No contribution proportional to the gravitational
anomaly coefficient
is found in this case.\\

{\bf Discussion.}
In vacuum the anomaly appears on the level of three point functions. In the
presence of external sources for the energy momentum tensor
and the currents this is conveniently expressed through \cite{GravAnom}
\begin{equation}
 \nabla_\mu J_A^\mu= \epsilon^{\mu\nu\rho\lambda}\left( \frac{d_{ABC}}{32\pi^2} 
F^B_{\mu\nu} F^C_{\rho\lambda} + \frac{b_A}{768\pi^2} 
\cR^\alpha\,_{\beta\mu\nu}
\cR^\beta\,_{\alpha\rho\lambda}\right) \,.\label{eq:anomaly}
\end{equation}
The anomaly coefficients are defined by
\begin{eqnarray}
d_{ABC} &=& \frac 1 2 [ \tr( T_A \{ T_B, T_C\} )_R -  \tr( T_A \{ T_B, T_C\} )_L
]\,, \label{eq:chiralcoeff}\\
b_A &=& \tr (T_A)_R - \tr (T_A)_L\label{eq:gravcoeff} \,,
\end{eqnarray}
where the subscripts $R,L$ stand for the contributions of right-handed and
left-handed fermions. 

We have computed the magnetic and vortical conductivity
at weak coupling and we find contributions that are proportional to
the anomaly coefficients (\ref{eq:chiralcoeff}) and
(\ref{eq:gravcoeff}). Non-zero values of these coefficients are a
necessary and sufficient condition for the presence of anomalies \cite{sufficient}.
Therefore the non-vanishing values of the transport coefficients
(\ref{eq:sigmaB}) and (\ref{eq:sigmaV}) have to be
attributed to the presence of chiral and gravitational anomalies.

Since the gravitational anomaly is fourth order in derivatives it
is a bit surprising to find it contributing to first order transport
coefficients. One possible intuitive explanation
one could think of is that the gravitational field in the presence of
matter gives rise to a fluid velocity $u^\mu$ e.g. through frame
dragging effects and that this might effectively reduce the number of
derivatives that enter in the hydrodynamic expansion.

The holographic calculation in
AdS/CFT~\cite{Amado:2011zx} did not show a
contribution proportional to $T^2$. This not surprising since only a
holographic gauge Chern-Simons term was included.
Holographic modeling of the gravitational anomaly calls however also for
inclusion of a mixed gauge-gravitational Chern-Simons term of the form 
$A\wedge R\wedge R$ \cite{progress}.

We find  a non-vanishing vortical conductivity proportional to $\sim T^2$ even
in an uncharged fluid.   In ~\cite{Neiman:2010zi} similar terms in the vortical
conductivities have been argued for as undetermined integration constants
without any relation to the gravitational anomaly.

It is also interesting to specialize our results to the case of one vector and
one axial current with chemical potentials $\mu_R = \mu+\mu_A$, $\mu_L= \mu
-\mu_A$, charges $q^R_{V,A}=(1,1)$ and $q^L_{V,A}=(1,-1)$ for one right-handed
and one left-handed fermion. We find (for a vector magnetic field)
\begin{eqnarray}
\sigma^\cB_{VV} = \frac{\mu_A}{2\pi^2}  &,&  \sigma^\cB_{AV} =
\frac{\mu}{2\pi^2} \,, \nonumber\\ \label{eq:axialcvc}
\sigma^\cV_{V} = \frac{\mu\mu_A}{2\pi^2} &,& \sigma^\cV_{A} =
\frac{\mu^2+\mu_A^2}{4\pi^2}+ \frac{T^2}{12}   \,.
\end{eqnarray}
Here $\sigma_{VV}^\cB$ is the chiral magnetic conductivity
\cite{Kharzeev:2009pj},  $\sigma_{AV}^\cB$ describes the generation of an axial
current due to a vector magnetic
field \cite{Son:2004tq}, $\sigma^\cV_{V}$ is the vector vortical conductivity,
$\sigma_{A}^\cV$ is the axial vortical conductivity and the only one 
sensitive to the gravitational anomaly.

In \cite{KerenZur:2010zw} enhanced production of high spin hadrons (especially $\Omega^-$ baryons) perpendicular to the reaction plane in heavy ion collisions 
has been proposed as an observational signature for  the chiral separation effect. Three sources of chiral separation have been identified: the anomaly in vacuum, the magnetic and the vortical conductivities of the axial current $J_A^\mu$. Of these the contribution of the vortical effect was judged to be subleading 
by a relative factor of $10^{-4}$. The~$T^2$ term in (\ref{eq:axialcvc}) leads however to a significant enhancement. If we take $\mu$ to be the baryon 
chemical potential $\mu\approx 10$ MeV, neglect $\mu_A$ as in \cite{KerenZur:2010zw} and take a typical RHIC temperature of $T=350$ MeV, we see that 
the temperature enhances the axial chiral vortical conductivity by a factor of the order of $10^4$. We expect the enhancement at the LHC to be even 
higher due to the higher temperature. 

Beyond applications to heavy ion collisions leading to charge and
chiral separation effects \cite{KerenZur:2010zw,Kharzeev:2010gr} it is
tempting to speculate that the new terms in the chiral vortical
conductivity might play a role in the early universe.  Indeed it has
been suggested before that the gravitational anomaly might give rise
to Lepton number generation, e.g. in \cite{Alexander:2004us}. The
lepton number separation due to the gravitational anomaly could
contribute to generate regions with non-vanishing lepton number. \\

\acknowledgements{
We thank D. Kharzeev, A. Rebhan and A. Schmitt for email conversation.
We also would like to thank ESI Vienna for hospitality during the
workshop on AdS Holography and Quark Gluon Plasma. This work has been
supported by Plan Nacional de Altas Energ\'{\i}as FPA2009-07908, CPAN
(CSD2007-00042), Comunidad de Madrid HEP-HACOS S2009/ESP-1473. }

\end{document}